\documentclass[aps,pre,showpacs,twocolumn]{revtex4}

\usepackage{graphicx}
\DeclareGraphicsExtensions{.png,.jpg,.eps}
\usepackage{amssymb}
\usepackage{braket}
\usepackage{xcolor}
\usepackage{amsmath}
\usepackage{color}
\usepackage{pbox}

\begin{document}

\title{Understanding the propagation of excitations in
quantum spin chains with different kind of interactions }

\author{Alejandro Ferr\'on$^{(1)}$, Pablo Serra$^{(2)}$ and
Omar Osenda$^{(2)}$}

\affiliation{(1) Instituto de Modelado e Innovaci\'on Tecnol\'ogica
(CONICET-UNNE) and
Facultad de Ciencias Exactas, Naturales y Agrimensura, Universidad Nacional
del Nordeste, Avenida Libertad 5400, W3404AAS Corrientes, Argentina.\\
(2) Instituto de F\'sica Enrique Gaviola, Universidad Nacional de C\'ordoba
, CONICET, Facultad de Matem\'atica, Astronom\'ia, F\'isica y Computaci\'on,
Av. Medina Allende s/n, Ciudad Universitaria, CP:X5000HUA C\'ordoba, Argentina
}

\date{\today}

\begin{abstract}

The dynamical behaviour of the quantum state of different quantum spin chains,
with designed site dependent interaction strengths, is analyzed when the
initial state belongs to the one excitation subspace. It is shown that the
inhomogeneous chains are able to transfer excitations with near perfect
fidelity. This behaviour is found for two very different spin chain
Hamiltonians. The first one is the ferromagnetic Heisenberg Hamiltonian with
nearest neighbor interactions, the second one describes a chain with long
range anisotropic interactions which are ferromagnetic in the $z$ direction and
antiferromagnetic
in the $(x,y)$ plane. It is shown that both designed chains have in common  a
partially
ordered spectrum and well localized eigenvectors. This physical trait unifies
the description of both kind of systems.

\end{abstract}

\maketitle

\section{introduction}

The dynamical evolution on the one excitation subspace has been studied in a
number of context ranging from the propagation of correlations in disordered
quantum
spin chains \cite{Burrell2007} to the transfer of quantum states
\cite{Bose2003,Bose-review,Nikolopoulos2015}.

When the Hamiltonian of the spin chain is ferromagnetic the link between the
propagation of a single excitation and the transfer fidelity of arbitrary
one-spin states is direct \cite{Bose-review}, since the fidelity of the
transfer process can be
written rather easily in terms of the probability that an excitation prepared in
one extreme of the chain will be tranferred to the other extreme.

The availability of numerous physical systems where it is possible to
actually implement the transfer protocols
\cite{Kandel2021,Martins2017,Kostak2007,quantum-dot-chain,Li2018,
nuclear-spin-chain,Loft2011,Banchi2011prl,Chapman2016,Kandel2019,Baum2021}
has driven
the study of different
control strategies to achieve the propagation of excitations in a ballistic
way, {\em i.e.} the propagation of a spin wave packet that propagates at
constant velocity along the chain. Nevertheless, this simple physical
behaviour can not be achieved easily in most quantum systems or in the spin
chains that model its behaviour.

Much progress has been made using time-dependent control techniques to achieve
near perfect quantum state transfer (QST)
\cite{Wang2016,Burgarth2010,Yang2010,Zhang2016,Farooq2015,Coden2021}, but the
conclusions of those studies
are, at some extent, particulars and do not inform about general physical
traits that lead to the design of general simple pulses that are applicable to
different systems. This is not necessarily a bad scenario, most spin chains are
controllable \cite{Jurdjevic1972,Burgarth2009,Wang2016,Ramakrishna1995} with a
reduced number of control functions (or a reduced number of
actuators) and the transmission of information is doable at times that scale
linearly or quadratically with the chain length $N$
\cite{Wang2016,Burgarth2010,Yang2010,Zhang2016,Farooq2015}.

More recently there is a renewed interest in the QST problem without
any external forcing since the fabrication techniques of microscopic quantum
systems allow to build systems with tailored interactions with
time-independent but site-dependent variable strength. Modulating the strength
of the interaction in order to achieve perfect or near perfect QST with
autonomous time evolution instead of a forced evolution has been the subject of
many recent works \cite{Zwick3,Banchi2010,Banchi2011}.

For  Heisenberg chains it is known that perfect QST transfer is not
achievable even changing the strength of the exchange
couplings \cite{Kay2019,Kay2010,Banchi2017,vanBommel2010}. For more
complicate models this is not even know, in many cases the analytical study of
the system is too cumbersome to reach any conclusion. Having this in mind, the
problem of QST can be cast as an optimization problem, in which the probability
that an excitation will be transferred between the extremes of a spin chain is
the cost function to be maximized, and the set of possible strength of the
interactions  the parameter space, where the optimal values of the interactions
must be found \cite{Serra2022,Zhang2018}.

Using a Globlal Optimization method guarantees that some good maximum value
will be found \cite{ssk97,sskb97}. We apply the pivot method to two models, the
Heisenberg
Hamiltonian and an anisotropic dipolar Hamiltonian, which has been studied as
an approximation to the Hamiltonian of cold atoms in optical traps
\cite{Hauke2010,Porras2004,Hung2016}.
Besides
showing that, effectively, the optimization method finds interaction strengths
compatible with near perfect QST we show that the optimal strengths have a
simple physical interpretation, that unifies the description of the time
evolution of the quantum state in long and short range models, irrespective of
their isotropic or anisotropic character. Our results also show that the
consequent dynamical evolution is smoother than the observed for homogeneous
chains whose time-evolution is marred by dynamical localization effects.

The paper is organized as follows, in Section~\ref{sec:models-and-methods} the
details of the pivot method, and how to apply it to the short and long range
models, are presented. In Section~\ref{sec:variable-number} the results
obtained for the propagation of excitations optimizing a variable number of
parameters that define the spin chains are presented, while the results
concerning the general properties of the spectrum and eigenvalues of spin
chains that show near perfect quantum state transfer are presented in
Section~\ref{sec:ordered-spectrum}. The behaviour of the inverse participation
ratio \cite{Giraud2007,Zwick2011} as a dynamical quantity is studied in
Section~\ref{sec:dynamical-localization}, in particular the study focuses in
the conditions that ensure a smooth, fast and efficient transfer of quantum
states. Finally, in Section~\ref{sec:conclusions} we discuss our results and
our conclusions are presented.

\section{models and method}\label{sec:models-and-methods}

We will focus our study in two well-known quantum spin chain Hamiltonians, the
isotropic short-range Heisenberg Hamiltonian

\begin{equation}
\label{eq:Heisenberg-Hamiltonian}
H_{short} = - \sum_{i=1}^{N-1} J_i \left( \sigma_i^x  \sigma_{i+1}^x +
\sigma_i^y  \sigma_{i+1}^y + \sigma_i^z  \sigma_{i+1}^z\right),
\end{equation}

\noindent where $J_i>0$, and the anisotropic long-range dipolar Hamiltonian

\begin{equation}
\label{eq:long-range-Hamiltonan}
H_{long} =  \sum_{i< j}^N \frac{J}{|x_i-x_j|^3}
\left(c_x \sigma_i^x  \sigma_{j}^x + c_y \sigma_i^y  \sigma_{j}^y +c_z
\sigma_i^z  \sigma_{j}^z\right),
\end{equation}

\noindent where $J>0$, $c_x=c_y=1$ and $c_z=-2$, {\em i.e.} the distance
dependent
exchange couplings in the plane are antiferromagnetic while in the $z$ direction
the exchange couplings are ferromagnetic. This last Hamiltonian can be found as
the weak limit of the Hamiltonian of a chain of cold atoms. In the following we
will call a particular set of $N-1$ exchanges coefficients $J_i$ an Exchange
Coefficients Distribution (ECD). In what follows we always consider {\em
centro-symmetric} chains, {\em i.e.} $J_i=J_{N-i}$, while for the long range
model  $|x_i - x_j| = |x_{N-i+1}-x_{N-j+1}|$.

The transfer properties of both Hamiltonians have been studied extensively when
the chains are homogeneous, {\em i.e.} all the $J_i$ coefficients  in
Eq.~\eqref{eq:Heisenberg-Hamiltonian} are equal, $J_i=J_h, \forall i$, and
$|x_i - x_{i+1}| = d > 0, \forall i$ in
Eq.~\eqref{eq:long-range-Hamiltonan}.
In particular, for the Heisenberg Hamiltonian it is known that the homogeneous
chain is quite poor as a transfer channel when the simplest transfer protocol is
implemented.

Both
Hamiltonians commute with the total
magnetization in the $z-$direction

\begin{equation}
 \left[ H, \sum_i \sigma_i^z \right] = 0,
\end{equation}

\noindent so both Hamiltonians can be diagonalized in subspaces with fixed
number
of excitations, {\em i.e.} in subspaces with a given number of spins up. It is
customary to use the computational basis, where for a single spin $|0\rangle
=|\downarrow\rangle$ and $|1\rangle =|\uparrow\rangle$, so
$|\mathbf{0}\rangle =
|0000\ldots 0\rangle$ is the state with zero spins up of the whole chain, and

\begin{equation}
 \sum_{i=1}^N \sigma_i^z |\mathbf{0}\rangle = -N |\mathbf{0}\rangle.
\end{equation}

The $N$ states with a single spin up are denoted as follows
\begin{equation}
|\mathbf{1} \rangle = |10\ldots 0\rangle, |\mathbf{2} \rangle =
|010\ldots0\rangle,\ldots , |\mathbf{N} \rangle = |00\ldots 1\rangle.
\end{equation}

\noindent {\em i.e.} the state $|\mathbf{j} \rangle $ is the state of the chain
with only the $j$-th spin up. The one-excitation basis plays a fundamental role
in the simplest transfer protocol. In this protocol the initial state of the
chain, $|\Psi(0)\rangle$, is prepared as

\begin{equation}\label{eq:initial-chain-product}
|\Psi(0)\rangle =|\psi(0)\rangle \otimes |\mathbf{0}\rangle_{N-1} ,
\end{equation}

\noindent where $|\psi(0)\rangle = \alpha |0\rangle +\beta |1\rangle $, is an
arbitrary
one-spin pure state, with $\alpha$ and $\beta$ complex constants such that
$|\alpha|^2 + |\beta|^2 =1$, and $|\mathbf{0}\rangle_{N-1}$ is the state
without excitations of a chain with $N-1$ spins. The state in
Eq.~\eqref{eq:initial-chain-product} can be rewritten as

\begin{equation}
 \label{eq:initial-state}
|\Psi(0)\rangle = \alpha |\mathbf{0}\rangle + \beta |\mathbf{1} \rangle .
\end{equation}
Using the time evolution operator
\begin{equation}
\label{eq:time-evolution}
U(t) = \exp{(-i H t)},
\end{equation}
the state of the chain at time $t$ can be obtained as
\begin{equation}
 \label{eq:time-dependent-state}
 |\Psi(t)\rangle = U(t) |\Psi(0)\rangle = \alpha U(t) |\mathbf{0}\rangle +
\beta U(t) |\mathbf{1}\rangle .
\end{equation}
It is clear that the state transfer is perfect when for some time $T_{per}$ the
state
of the chain is given by
\begin{equation}
\label{eq:perfect-transfer}
|\Psi(T_{per})\rangle = e^{i\theta}\left(\alpha |\mathbf{0}\rangle + \beta
|\mathbf{N}\rangle\right),
\end{equation}
where $\theta$ is some real arbitrary phase. So, the transfer fidelity as a
function of time is calculated as

\begin{equation}\label{eq:first-fidelity}
 f_{\alpha,\beta}(t) = |\alpha \langle \Psi(t)|\mathbf{0}\rangle + \beta
\langle \Psi(t)
|\mathbf{N}\rangle|^2 .
\end{equation}

The fidelity in Eq.~\eqref{eq:first-fidelity} depends on the particular initial
state, Eq.~\eqref{eq:initial-state}, chosen to be transferred. Usually, to
quantify the quality of the state transfer protocol, it is introduced the
average of  $f_{\alpha,\beta}(t) $ over all the possible values of $\alpha$ and
$\beta$. It can be shown that this averaged fidelity, $F(t)$, can be written as

\begin{equation}\label{eq:fidelity}
 F(t) = \frac{\sqrt{|P(t)|}}{3} + \frac{P(t)}{6}+\frac12 ,
\end{equation}

\noindent where

\begin{equation}\label{eq:population}
P(t) = |\langle \mathbf{1}|U(t)|\mathbf{N}\rangle|^2
\end{equation}

\noindent is
called the
 transferred population between the first and last sites of the
chain,  and we assume that $\cos(\gamma)=1$, where
$\gamma=arg(f_{\alpha,\beta}(t))$,  as in Ref.~\cite{Bose2003}. The
pretty good transfer occurs when for some $t_{\epsilon}$

\begin{equation}
 P(t_{\epsilon}) = 1- \epsilon, \qquad \forall \epsilon>0 .
\end{equation}
\\

The idea behind our method is quite simple and can be stated as follows: it is
simpler to use inhomogeneous chains with site-dependent exchange coefficients
tailored to achieve a high probability of transfer which is carried out by
the autonomous dynamics of the system than to control the dynamics using
time-dependent pulses. The tailoring is made using a Global optimization method
that finds a good maximum value for $P(T)$ irrespective of the Hamiltonian or
the
arrival time $T$.

\begin{figure}[hbt] 
\includegraphics[width=0.95\linewidth]{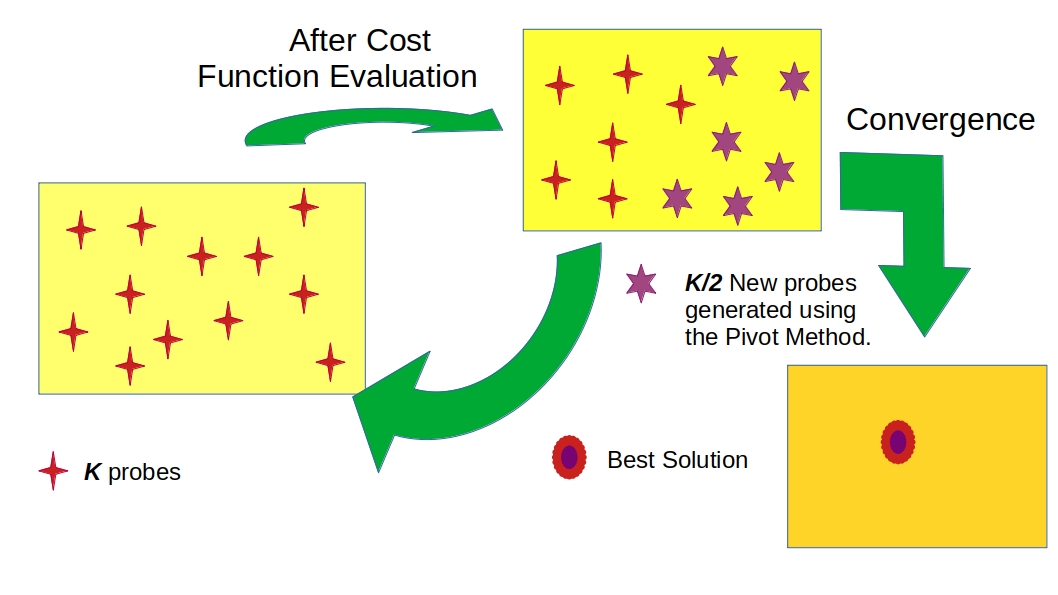}
\caption{The cartoon depicts how the pivot  method works. Starting from $K$
randomly chosen probes (the cross-like red dots) the cost function is evaluated
for each one of them. Then, the $K/2$ probes whit the largest values of the
cost function are kept (those with the lowest values are discarded) and the
method {\em pivots} over them to generate other $K/2$ new ones. The cost
function is evaluated and the convergence checked. If the convergence criterion
is not satisfied new rounds of discarding and pivoting are repeated until
convergence is achieved.
}\label{fig:cartoon1}
\end{figure}

In Reference~\cite{Serra2022} we used a Global optimization method to obtain
chains that showed near perfect QST so the detailed description of the method
can be consulted there. Here we describe briefly how the method proceeds for
the short range model, but the procedure is exactly the same for any
one-dimensional chain.

The search for an optimal value of a given cost function, that depends on a
number of variables, starts from a set of probes which are randomly generated
in a hypercube whose side length is chosen to contain the possible values  of
the variables that determine the value of the cost function. In the cartoon in
Figure~\ref{fig:cartoon1} the yellow rectangle depicts the hypercube and the
starry dots show the  set of probes. For the initial set of $K$ probes
the cost function is evaluated and the set of $K/2$ probes with the larger
values of the cost function are kept. Then the pivot method generates new $K/2$
probes. After this step convergence is checked and if the criterion is
satisfied the probe with the largest cost function value is selected. If the
convergence is not achieved  $K/2$ probes are discarded (those with the lower
values of the cost functions) and the pivot method is called again (for more
details see References~\cite{ssk97,sskb97}). This steps
are depicted in the cartoon, the green arrows showing the direction that the
algorithm takes until the convergence is achieved.

\begin{figure}[hbt] 
\includegraphics[width=0.95\linewidth]{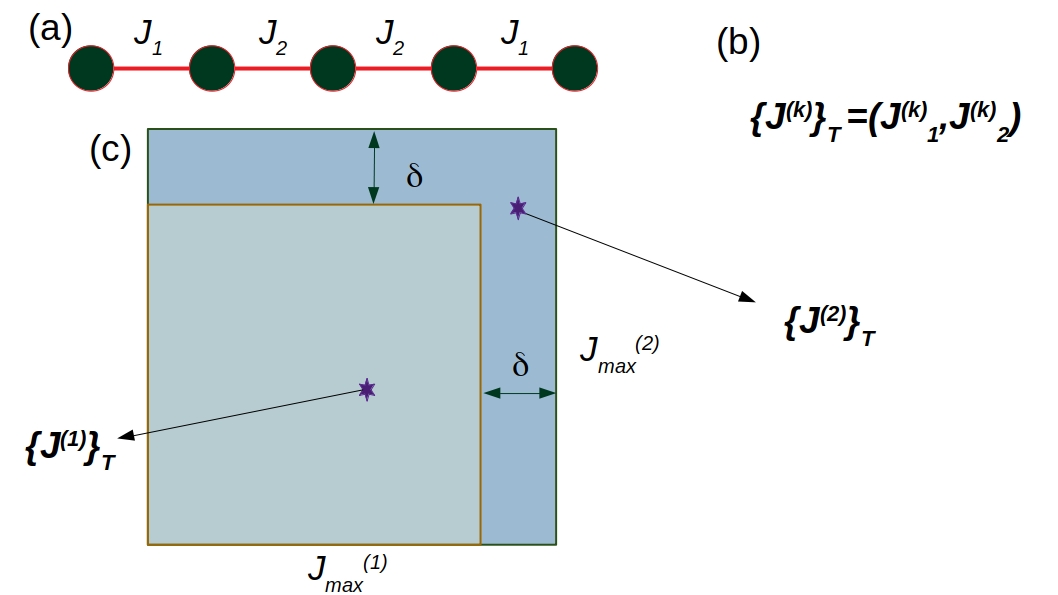}
\caption{The cartoon depicts the ingredients of the procedure designed to find
ECD's that can be used to achieve near perfect QST. a) A
centro-symmetric spin chain with five spins and only two different exchange
couplings $J_1$ and $J_2$. b) For the five spin chain a ECD that optimize the
cost function for an arrival time $T$ is obtained using  the pivot method . The
ECD is denoted by $\lbrace J^{(k)} \rbrace_T =(J^{(k)}_1,J^{(k)}_2)$, which is a
point in the hypercube with side length  $J_{max}^{(k)}$. c) The smaller
hypercube with side length $J_{max}^{(1)}$ and the second one smaller, with side
length $J_{max}^{(2)} = J_{max}^{(1)}+\delta$, are depicted using light blue and
blue squares. The corresponding ECD that optimize the cost function for each
hypercube is depicted as a starry dot inside the hypercube.
}\label{fig:cartoon2}
\end{figure}

It is clear that for problems of QST the cost function to be maximized is the
PT, $P(T)$, at some arrival time, $t=T$. For general problems the side length
of the
hypercube should be carefully chosen, since the volume of the hypercube is
$L^M$, where $M$ is the number of variables that determine the value of the
cost function. Because of this we proceed as follows, first we start using the
pivot method in a hypercube with a relatively small side length and applying
the pivot method a first ECD is determined, $\lbrace J^{(1)} \rbrace_T$, which
 maximizes the PT for an arrival time $T$. Then the side length of the
hypercube is increased and a second ECD is obtained,  $\lbrace J^{(2)}
\rbrace_T$. This procedure is depicted in the cartoon in
Figure~\ref{fig:cartoon2} for a centro-symmetric
chain with five spins that has only two different exchange couplings. The
initial hypercube is depicted with light blue square with side
length $J_{max}^{(1)}$ while the second hypercube with side
length $J_{max}^{(2)}$ is depicted as a darker blue square. The notation is
used to emphasize the succession of values obtained. Calling $P^{(k)}(T)$ to
the PT value obtained using the ECD $J_{max}^{(k)}$, it is fulfilled that

\begin{equation}
 P^{(1)}(T) \leq P^{(2)}(T) \leq P^{(3)}(T) \leq \ldots .
\end{equation}

\noindent In \cite{Serra2022} it is shown that for anisotropic and isotropic
Heisenberg Hamiltonians with only nearest neighbors interactions the side
length of the successive hypercubes is a decreasing function of the arrival
time and is bounded above by $N$.

It is worth to point out that for $k$ large enough the optimization method
converges to a given ``optimal'' ECD and PT, this value for the PT is the one
that is plotted in the figures of the next Sections. Nevertheless, sometimes it
is useful to study the behaviour of the ECD for smaller values of the PT to
understand how the spectral properties of a given Hamiltonian change with the
value of the target PT.

The procedure described above can be easily translated to other Hamiltonians.
For the long range dipolar Hamiltonian in
Eq.~\eqref{eq:long-range-Hamiltonan} we consider the distances between
nearest neighbors $\lbrace |x_i-x_{i+1}| \rbrace_{i=1}^{N-1}$ as the variables
to be explored by the pivot method. The cost function is again the PT at some
arrival time $T$.

\section{QST optimizing a variable number of exchange
coefficients} \label{sec:variable-number}

In Reference~\cite{Serra2022} it was shown that the optimization procedure
produces ECDs $\lbrace J^{(k)} \rbrace_T$, for $k$ large enough, that result in
near perfect QST when the Hamiltonian of the chain is of  the $XXZ$ type with
nearest neighbors interactions. In this Section we want to address two
related questions. The first one is: if only a few exchange coefficients (or
distances) can be optimized, due to some implementation limitation,  is it
possible to achieve near perfect or very good QST? The second one is: short
range and long range spin chains with only few quantities optimized behave in a
similar way?

\begin{figure}[hbt] 
\includegraphics[width=0.95\linewidth]{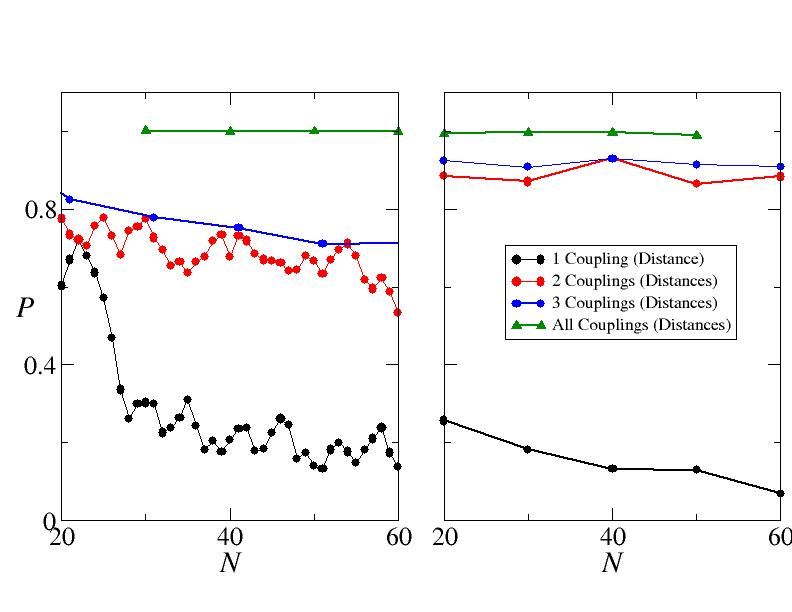}
\caption{The population transferred as a function of the chain length $N$.
Panel a) shows the results obtained for the short range model, panel b) shows
the results for the long range model. From bottom to top the curves are
obtained optimizing one (black dots and curve), two (red), three (blue) and all
(green) the exchange couplings coefficients.
}\label{fig:variable-number}
\end{figure}

Figure~\ref{fig:variable-number} summarizes our results about the two questions
stated above. The figure shows the population transferred,
for different chain lengths and optimizing (from bottom to top) one, two, three
or $N/2$ exchange couplings or distances, accordingly with the model. The
arrival time considered is $T=2N$. The left panel shows the results found for
the Heisenberg model, while the right one shows the results found for the
anisotropic long-range dipolar model. It is clear that the PT is an increasing
function of the number of exchange couplings optimized, which is to be
expected. Nevertheless, what is striking is the fact that the long range model
increases its efficiency as a quantum channel faster than the short range one.
Of course, if the $N/2$ exchange couplings or distances are optimized the GO
method provides
chains that show near perfect QST. Similar scenarios can be found for other
arrival times.

The results shown in Figure~\ref{fig:variable-number}, despite its usefulness,
do not provide an insight about the physical mechanisms that is responsible of
the increased efficiency of the chains with more exchange couplings optimized.
The next
Sections are devoted to understand these mechanisms.

\section{Eigenvector localization and partially ordered spectrum}
\label{sec:ordered-spectrum}

Following Kay \cite{Kay2019}, near perfect QST can be expected in Heisenberg
spin
chains whose spectrum satisfies $E_{i+1} - E_{i} \sim q_i \alpha$, where $q_i$
is
an odd natural number, {\em i.e.} successive energy values should differ in odd
multiples of a constant $\alpha$, approximately.

\begin{figure}[hbt] 
\includegraphics[width=0.95\linewidth]{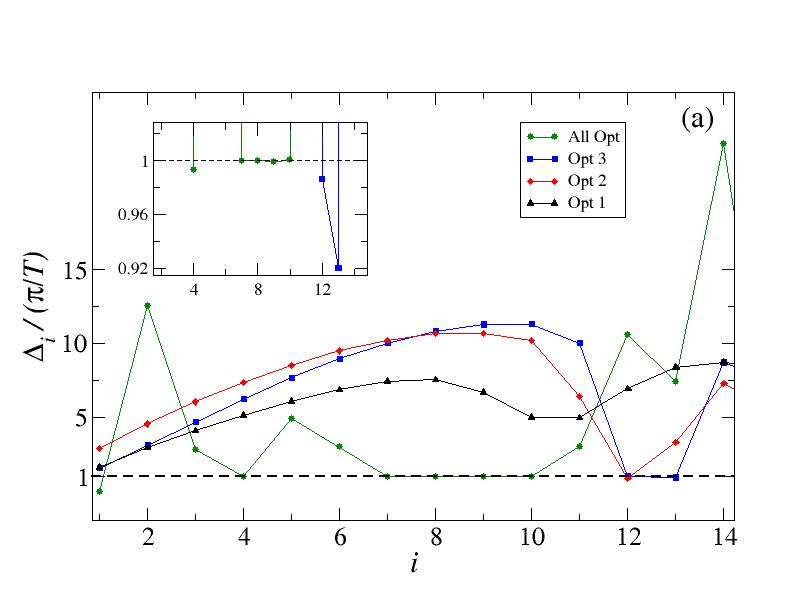}
\includegraphics[width=0.95\linewidth]{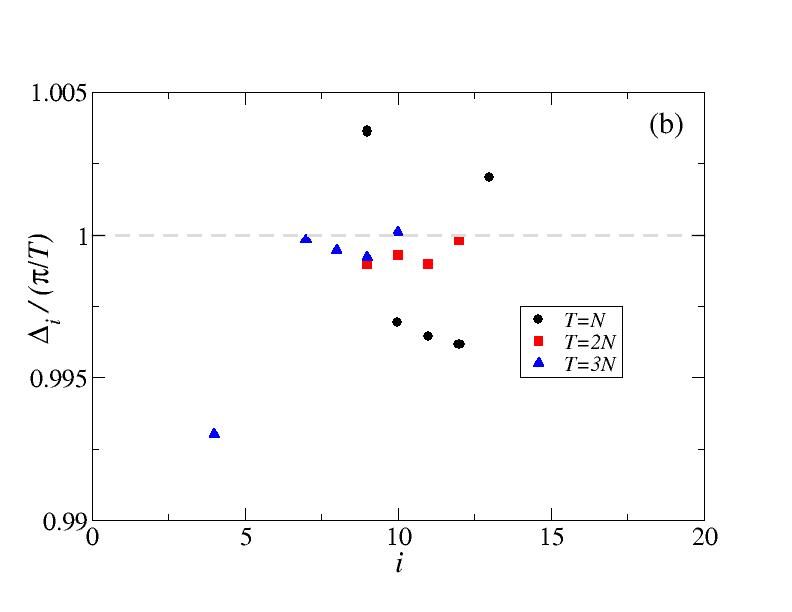}
\caption{The scales difference between successive eigenvalues, $\Delta_i$, for
the optimal exchange coupling distribution {\em vs} the index $i$, for a chain
with length $N=20$ and an arrival time $T=2N$. a) As in
Figure~\ref{fig:variable-number}, the data corresponding to the results obtained
optimizing one, two, three or all the exchange couplings are shown using black,
red, blue and
green dots, respectively. Panel b) shows a detailed view of the zone near
$\Delta_i=1$, for a chain with length $N=40$ spins and different arrival times.
}\label{fig:ordered}
\end{figure}

Figure~\ref{fig:ordered}a) shows the difference between successive
eigenvalues, $ \Delta_i = E_{i+1} - E_{i}$ for a chain with $N=20$ spins {\em
vs} the eigenvalue index $i$, for chains with different number of distances
optimized. The arrival time considered is $T=2N=40$.  The distance $\Delta_i$
is scaled with $\pi/T$. It is clear that the number of eigenvalues that
satisfies $\Delta_i/(\pi/T)\sim q_i$, increases as a function of the
number of distances optimized. The inset shows that $\Delta_i/(\pi/T)\sim 1$ is
satisfied with a remarkable precision when $N/2$ distances are optimized. We
called this scenario a partially ordered spectrum.

The
time dependent state of the chain can be written as

\begin{equation}\label{eq:psi-decomposed}
\psi(t) = \sum_{\tilde{i}} \chi_{\tilde{i}} \exp{\left( -i E_{\tilde{i}}
t\right)}
+ \sum_{i\neq\tilde{i}} \chi_{i} \exp{\left(-i E_{i} t\right)} ,
\end{equation}

\noindent where the index $\tilde{i}$ runs over the eigenvalues that satisfies
$\Delta_{\tilde{i}}/(\pi/T)\sim q_i$, then at $t=T$ the terms on the first sum
will
interfere constructively. Near perfect QST occurs at arrival time $T$ when a
number of terms in Eq.~\eqref{eq:psi-decomposed} interfere constructively
to
produce $|\langle \psi(T)| N\rangle |^2 \sim 1$.
The relatively reduced number of eigenvalues that
satisfies approximately the constructive interference condition

\begin{equation} \label{eq:constructive interference}
 \Delta_i/(\pi/T)= q_i
\end{equation}

\noindent indicates that to obtain near perfect QST it is necessary to get
$\chi_{i}\sim 0$ for $i\neq\tilde{i}$. Note that, as is shown in
Figure~\ref{fig:ordered}b), the number of eigenvalues that
approximately satisfies the constructive interference condition does not change
very much when the arrival time is increased, but those near unity become even
closer to it.

\begin{figure}[hbt] 
\includegraphics[width=0.95\linewidth]{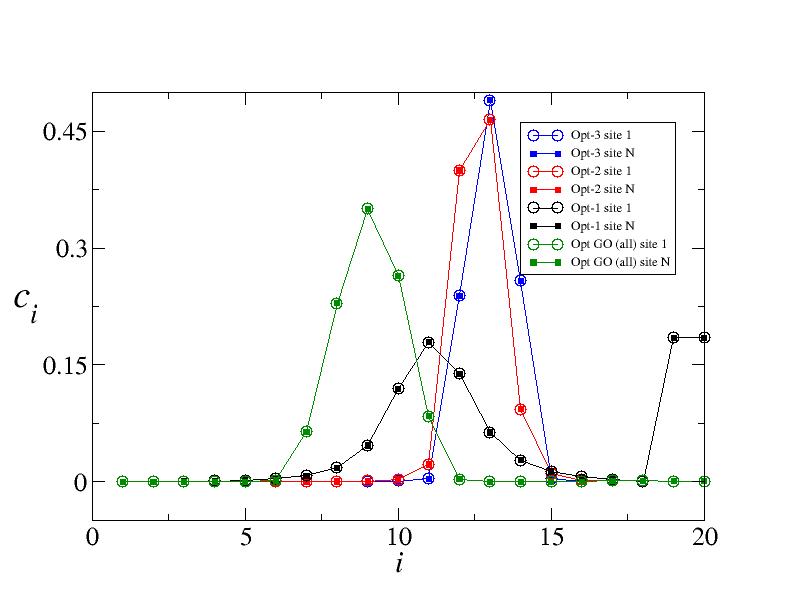}
\caption{The figure shows the coefficients $c_i=|\langle v_i|
\mathbf{1}\rangle|^ 2$, the square modulus of the scalar product between the
$i$-th eigenvector, $|v_i\rangle$, and $| \mathbf{1}\rangle$, {\em vs} $i$.
Again, the results corresponding to the optimization of one, two, three or all
the exchange couplings are shown using black, red, blue and green dots,
respectively.
}\label{fig:c-sub-one}
\end{figure}

Figure~\ref{fig:c-sub-one} shows the coefficientes $c_i = |\langle
v_i|\mathbf{1}\rangle|^2$ {\em vs} $i$, where the vectors $|v_i\rangle$ are
solutions of
\begin{equation}
 H ({J}_T) |v_i\rangle = E_i |v_i\rangle .
\end{equation}
Extended eigenstates have $c_i \sim 1/N$, where $N$ is the chain lenght.
Figure~\ref{fig:c-sub-one} shows that the eigenvectors corresponding to the
almost ordered eigenvalues are strongly localized at both extremes of the
quantum chain.

In a few words, almost near perfect QST is achievable with a reduced number of
eigenvalues satisfying condition \eqref{eq:constructive interference} as long
as
their eigenvectors are well localized
at the extremes of the spin chain. This scenario was at some extent, envisioned
as a necessary condition by Kay and others \cite{Kay2019,Banchi2017} a
requisite to achieve the so
called pretty good quantum state transfer, but for the isotropic Heisenberg
model. What our results indicate is that the pivot method finds precisely the
ECD that result in the almost ordered spectrum with well localized
eigenvectors that are required for the almost perfect QST to happen also for
the long range model in Eq.~\eqref{eq:long-range-Hamiltonan}.

The partially ordered spectrum with well localized
eigenvectors scenario is also present in the short range Heisenberg model,
Eq.~\eqref{eq:Heisenberg-Hamiltonian} so near perfect QST is also
achievable. Nevertheless there are some differences that are worth of
further analysis.

\begin{figure}[hbt] 
\includegraphics[width=0.95\linewidth]{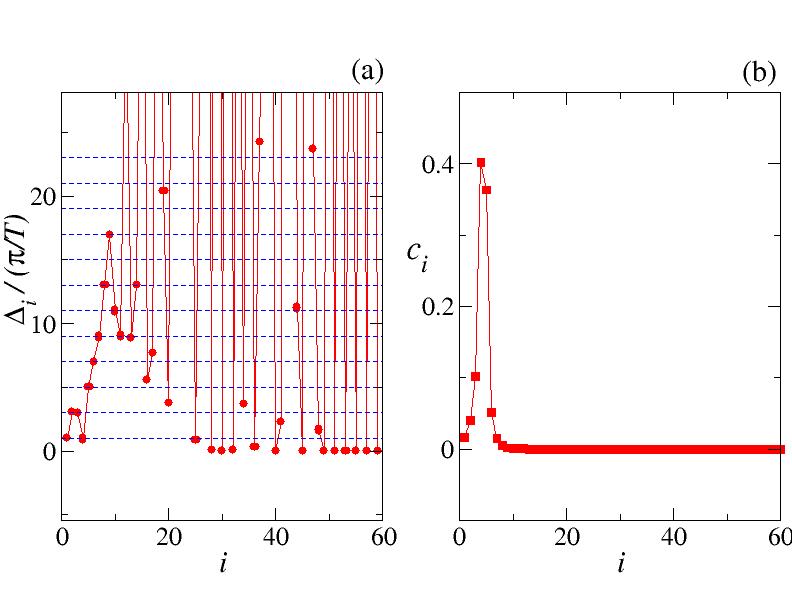}
\caption{The scaled difference between successive eigenvalues, $\Delta_i$, and
the $c_i$ coefficients, are shown in panels a) and b), respectively, both {\em
vs} the index $i$ for a chain with a short range Heisenberg Hamiltonian. The
chain has $N=60$ spins. In panel a) the partially ordered spectrum can be
clearly appreciated at the edge of the eigenvalues band, with values of
$\Delta_i$ close to odd naturals as large as 15 or 17. The data in panel b)
shows that only the eigenvectors corresponding to the partially ordered portion
of the spectrum are localized at the extremes of the chain.
}\label{fig:Heisenberg-one}
\end{figure}

Figure~\ref{fig:Heisenberg-one} shows a) $\Delta_i$ and b) $c_i$ both {\em vs}
the index $i$ for the Heisenberg Hamiltonian and a chain with $N=60$ spins.
There is a large number of successive eigenvalues that satisfy the condition in
Eq.~\eqref{fig:ordered}, anyway those that satisfy the condition with
$q_i=1$ are a minority. In the figure, values of $\Delta_i=15, 17$ can be
clearly
appreciated. The appearance of values of $\Delta_i$ is related to the different
propagation velocities that are characteristic of the Heisenberg model.
Nevertheless, Figure~\ref{fig:Heisenberg-one} b) shows again, that the
eigenvectors associated to the part of the spectrum that is partially ordered
are well localised at the extremes of the chain. Other feature that
distinguishes the spectra of the short and long range model is that in the
spectrum of the former there is a number of very small eigenvalues,
$E_i\sim 0$.

Since the optimization method is interrupted when the criterion of convergence
is fulfilled it is natural to wonder about the spectrum of Hamiltonians that
correspond to ECD obtained with sub-optimal convergence criteria, {\em i.e.}
when the PT is far from near perfect QST.

\begin{figure}[hbt] 
\includegraphics[width=0.95\linewidth]{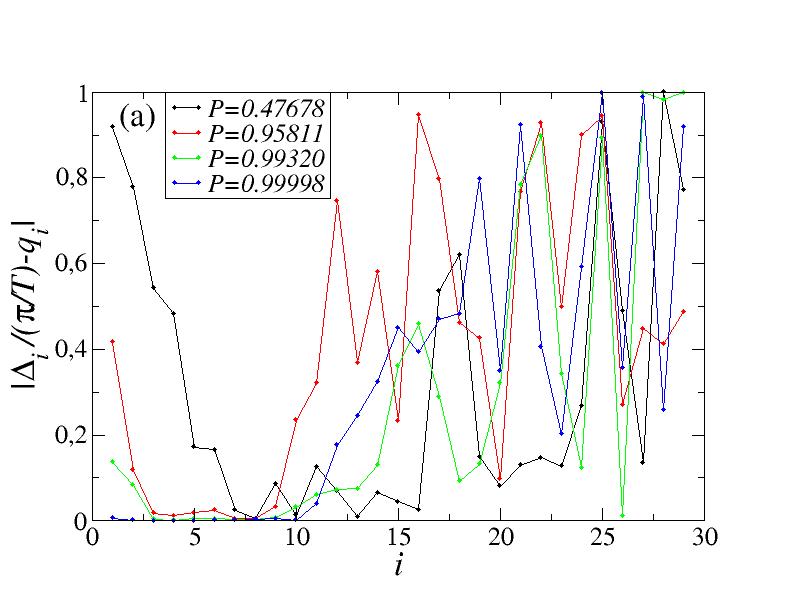}
\includegraphics[width=0.95\linewidth]{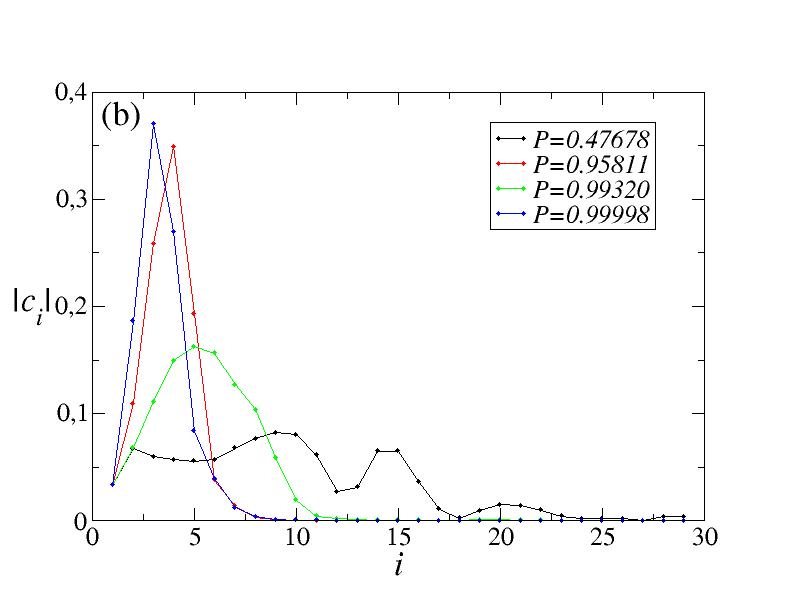}
\caption{The scaled difference between successive eigenvalues, $\Delta_i$
minus the closest odd integer, $q_i$, and
the $c_i$ coefficients, are shown in panels a) and b), respectively, both {\em
vs} the index $i$ for a chain with a short range Heisenberg Hamiltonian.
The
chain has $N=30$ spins, arrival time $T=N$ and the data is obtained
interrupting the optimization procedure at sub-optimal PT (see the text for more
details). The panels show the values obtained for PT values of $0.9998$,
$0.99320$, $0.95811$ and $0.47678$ using blue, red, green and black dots,
respectively. The lines are included as a guide to the eye. Panel a) shows how
the partially ordered portion of the spectrum increases when the PT value
becomes larger. The localization of the corresponding eigenvectors is shown in
panel b) and it can be appreciated how it becomes stronger for larger values of
the PT.
}\label{fig:Heisenberg-two}
\end{figure}

Figure~\ref{fig:Heisenberg-two} a) shows the distance $d_i$ which is defined as
the difference between $\Delta_i$ and its nearest odd natural for four
different values of the PT. The values were obtained allowing the optimization
of all the couplings. The data in the figure shows that the number of ordered
pair of eigenvalues is an increasing function of the transferred
probability at a given arrival time. Even for the largest PT achieved, that
differs from the unity in less than $2\times 10^{-4}$, only a third of the
eigenvalues are very close to an odd natural number, emphasizing that the
localization of the corresponding eigenvectors is also required to achieve near
perfect QST. The behaviour of the coefficient that measures how localized are
the eigenvectors at the extremes of the chain can be observed in
Figure~\ref{fig:Heisenberg-two} b).

\section{Preventing dynamical localization} \label{sec:dynamical-localization}

Besides the  effect of the optimization of the exchange couplings over
the spectrum and eigenvectors, which results in very large values for the PT,
there are other dynamical effects that take place during the transmission. A
quantity that allows to qualitatively understand the changes in the dynamical
behaviour of the chain state is the inverse participation ratio (IPR)

\begin{equation}
 L(t) = \frac{1}{\sum_i |\chi_i(t)|^4}.
\end{equation}

The IPR has been used to analyze the dynamical behaviour of many different
systems, ordered, chaotic or disordered, since it is able to detect how an
initially localized state spreads, or not, over the whole system.
The dynamical evolution of an one excitation propagating on Heisenberg-like
chains are characterized by rapid fluctuations in the value of the IPR since
the dynamics reflects the peculiar distribution of eigenvalues on the spectrum.
But, as was shown in Section~\ref{sec:variable-number} the optimization
procedure results in chains with partially ordered spectrum so the behaviour
of the IPR should reflect the emerging ordering and the appearance of well
localized eigenvectors.

\begin{figure}[hbt] 
\includegraphics[width=0.95\linewidth]{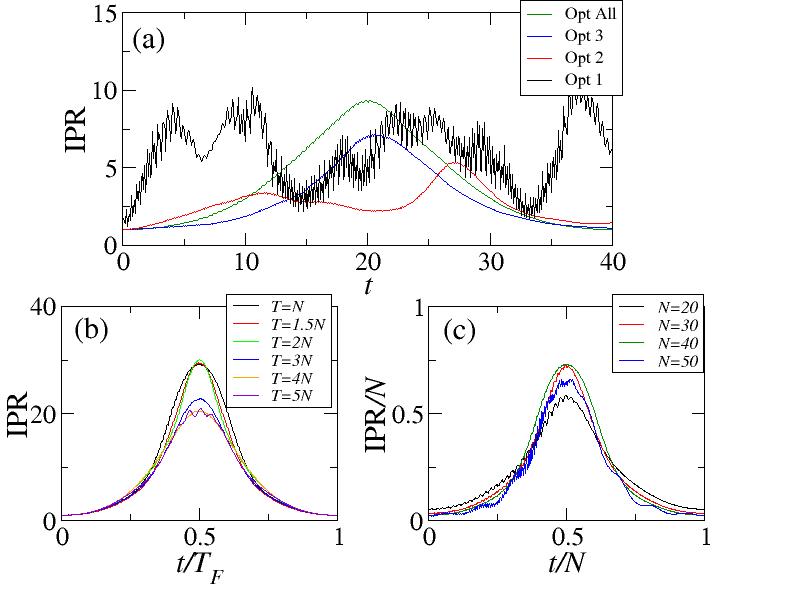}
\caption{The IPR as a function of time. Panel a) shows the results calculated
when one, two, three or all the exchange coefficients are optimized as
black, red, blue and green curves, respectively. The results correspond to a
chain with $N= 20$,  arrival time of $T= 40$. The chain considered is described
by
the long range Hamiltonian.  Panel b) shows the behaviour of the IPR for a
chain with $N=40$ spins but for different arrival times, each curve is plotted
{\em vs} the scaled time $t/T$. Panel c) shows the IPR {\em vs} the scaled
time $t/N$for chains of different
length but with their Hamiltonians optimized for the same arrival time $T=40$.
}\label{fig:ipr}
\end{figure}

Figure~\ref{fig:ipr}a) shows the time evolution of the IPR ,
calculated for a
chain of $N=20$ spins that has the long range dipolar Hamiltonian. The  curves
are obtained optimizing different number of distances. The effect of the
optimization and the progressive building of the partially ordered spectrum is
easily appreciated, since the IPR goes from a fast changing functions with
rapid oscillations to a smooth function when the number of quantities optimized
is increased.

On the other hand, Figure~\ref{fig:ipr} b) and c) show that in the ``optimized
regime'' the dynamical evolution of the excitation propagating in the chain is
pretty much the same. Panel b) compares the time evolution of the IPR for a
chain with $N=40$ spins and different arrival times, while panel c) shows the
same quantity for chains of different length and the same arrival time.

A similar analysis can be carried out with respect to the Heisenberg
Hamiltonian. The conclusions are quite similar, and they can be found in
Appendix~\ref{ap:heisenberg}, together with a figure showing the behaviour of
the IPR for this case.

\section{Discussion and Conclusions}\label{sec:conclusions}

The role played by localized states in QST settings is well known in the
context of XX Hamiltonians, where the regimes known as {\em border controlled
QST} \cite{Zwick3} and {\em optimal dynamics} \cite{Banchi2010} exploit that
changing the values
of just one or two  exchange couplings at the extremes of centro-symmetric
chains results in
regimes where near perfect o very good QST is achievable in otherwise
homogeneous chains. More recently, Palaiodimopoulos {\em et al} showed that the
localized eigenstates  lying at the centre of the eigenvalues band are
crucial to obtain fast and robust QST in a topological chain
\cite{Palaiodimopoulos2021}, although their protocol involved precise
time-dependet control of the exchange coupligs at the extremes of a SSH chain,
$J_1(t)$ and
$J_{N-1}(t)$. As our results show, looking for localized eigenstates
located at any region of the spectrum seems to be a requisite to attain
near perfect QST in arbitrary spin chains, as long as these eigenstates
correspond to the partially ordered portion of the spectrum. Of course, the
intricacies of the Bethe ansatz prevents that this traits can be singled out in
the Heisenberg Hamiltonian. The perspectives are even worse when the
anisotropic long range dipolar Hamiltonian is considered.

As our results suggest, it is harder to find the regime of smooth
propagation for the Heisenberg Hamiltonian than for the long range Hamiltonian.
This is another manifestation of the ``disorderly'' way in which the
excitations propagate in Heisenberg chains.

It is striking the reduced number of eigenvalues fulfilling
Eq.~\eqref{eq:constructive interference} that are really necessary to achieve
near perfect QST, as long as their eigenvectors are well localized at the
extremes of the chain. Since most studies showing the existence of the pretty
good QST focus in the requirements that the eigenvalues should satisfy, it is
worth to think if a change of strategy adding this fact could lead to new
advances on the design of Hamiltonians that can be implemented in experimental
settings.

With respect to the last point commented in the paragraph above, there are at
least to possible lines of work. The first one is to extend the ideas in
Reference~\cite{Kay2006}, where A. Kay discusses how to obtain near perfect QST
in chains with interactions beyond nearest neighbors. The second one requires
the implementation of algorithms characteristic of inverse eigenvalue problems,
that allow to construct a matrix with  prescribed eigenpairs. Work along this
lines is in progress

{\em Acknowledgments}
The authors acknowledge partial financial  support from
CONICET (PIP11220150100327, PUE22920170100089CO). O.O and P.S. acknowledges
partial
financial support from CONICET and SECYT-UNC. A.F. thanks the hospitality of
the FAMAF, where a large part of this manuscript was discussed and planned.

\appendix*

\section{IPR behaviour for Heisenberg chains}\label{ap:heisenberg}

\begin{figure}[hbt] 
\includegraphics[width=0.95\linewidth]{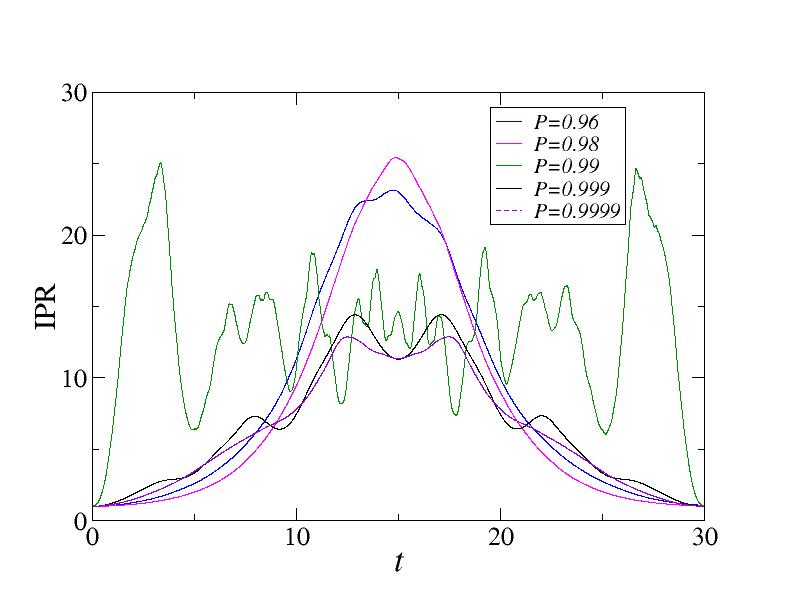}
\caption{The behaviour of the inverse
participation ratio, as a function of the time $t$, calculated using
sub-optimal ECD. The chain considered has $N=30$ spins and the optimal ECD
corresponds to an arrival time $T=N$.
}\label{fig:ipr-heisenberg}
\end{figure}

As has been said in the main text, the behaviour of the IPR as a function of
time for Heisenberg chain  is similar to the observed in the long range model
studied. Anyway, there are some qualitative differences that are interesting to
look at.
Figure~\ref{fig:ipr-heisenberg} shows the behaviour of the inverse
participation ratio, as a function of the time $t$, calculated using
sub-optimal ECD. The chain considered has $N=30$ spins and the optimal ECD
corresponds to an arrival time $T=N$. Note that, even for values as large as
$P=0.99$ the IPR shows rapid oscillations. Moreover, the dynamical regimes at
both sides of $P=0.99$ seem quite different, resulting in a dynamical evolution
with a single or double maximum for $P<0.99$ and $P>0.99$, respectively. For
smaller values of $P$ the ``bell-shape'' becomes increasingly distorted until
quite rapid oscillations spoil completely the transfer process (not shown).

\end{document}